\documentstyle[12pt]{article}

\newcommand{\newsection}{
\setcounter{equation}{0}
\section}
\newcommand{\tr}{\,{\rm tr}\,}
\def\e{{\,\rm e}\,}

\def\eop{\vspace*{\fill}\pagebreak}
\def\be{\begin{equation}}
\def\ee{\end{equation}}
\def\bea{\begin{eqnarray}}
\def\eea{\end{eqnarray}}

\def\t#1{\widetilde{#1}}

\def\EX{{\vec {EX}}}
\def\Bbb#1{{\bf {#1}}}
\def\ev#1{{\vec e}_{#1}}
\def\av#1{{\vec a}_{#1}}
\def\Tr{T_{red}}

\newcommand{\ie}{{\it i.e.}\ }

\renewcommand{\d}{{{\partial}}}

\title{{\bf \mbox{} \\$L$--functions in Scattering on $p$-adic Multiloop
Surfaces}
\vspace{.5cm}}
\author{{\bf L. Chekhov}\thanks{E--mail: \ chekhov@qft.mian.su}
\thanks{This work was supported in part by Russian F.F.I. grant
\#94-01-00285}\\
\date{ }
\vspace{.3cm} \\
{\it Steklov Mathematical Institute} \\
{\it Vavilov st.42, GSP-1, 117966 Moscow, Russia}}

\begin{document}

\maketitle

\vspace{-9.6cm}

\begin{flushright}
SMI--94--10 \\ March 29, 1994
\end{flushright}

\vspace{6.8cm}

\begin{abstract}
We study scattering processes on $p$-adic multiloop surfaces
represented as multiloop infinite graphs with total valence in each vertex
equal $p+1$. They all are spaces of the constant negative curvature
since they are quotients of the $p$-adic hyperbolic plane over
free acting discrete subgroup of $PGL(2, {\bf Q}_p)$. Releasing the
closed part of this graph containing all loops which is called
reduced graph $\Tr$ we can obtain $L$-function corresponding to this
closed graph.  For the total graph we introduce the notion of the
spherical functions being eigenfunctions of the Laplace operator
acting on the graph and consider $s$--wave scattering processes
therefore defining scattering matrices $c_i$. The number of
possibilities coincides with $|\Tr|$ --- the number of vertices of
the reduced graph. Taking the product over all $c_i$ we define the
total scattering matrix which appears to be essentially presented as
a ratio of $L$--functions:  $C\sim L(\alpha_+)/L(\alpha_-)$, where
the function $L$ itself depends only on the shape of $\Tr$ and not on the
initial infinite graph, and the only dependence of initial $p$ is contained
in arguments $\alpha_\pm$ defined by $p$ and eigenvalue $t$ of the
Laplacian. We also present a proof by H.Bass of the theorem
expressing $L$--functions on arbitrary finite graphs via determinants
of some local operators on these graphs.


\end{abstract}

\eop

\newsection{Introduction}
The purpose of this paper is to find an analogue of celebrated
Selberg trace formula \cite{Sel} for determinant of the Laplace
operator on Riemann surfaces of constant negative curvature for the
case of $p$-adic multiloop surfaces. Let us first summarize all known
(at least, to the author) facts:

{\bf 1.} Multiloop $p$-adic surfaces were introduced in \cite{M-S}.
They can be explicitly presented as a set of (infinite) graphs with a
finite number of primitive cycles, the number of them being $g$ --
the genus of the surface. Fixing the prime number $p$, the valences
of all vertices of the graph are the same -- $p+1$. Then one can
introduce a Laplace operator $\Delta$ acting on the space of
functions $C_0$ of the vertices of the graph, $\Delta
f_0=\sum_{neigh.}f_i-(p+1)f_0$, where the sum runs over all
neighbours of the point $x_0$, $f_i\equiv f(x_i)$. By the
construction we can release from the graph $X$ its ``closed'' part --
the reduced graph $\Tr$ containing all loops while valences of
vertices inside this graph can be arbitrary ($\le p+1$). In the paper
\cite{CMZ} the string theory for such graphs was developed and it was
demonstrated  that when calculating the corresponding amplitudes all
crucial ingredients of the string theory as prime forms, Schottky
groups, etc have their proper $p$-adic analogies. It is worth to note
that due to Schottky uniformization all these $p$-adic surfaces are
surfaces of the constant negative curvature.

{\bf 2.} For ordinary closed Riemann surfaces of constant (negative)
curvature the Selberg trace formula is valid \cite{Sel} that
establish an explicit relation between determinant of the Laplace
operator and zeta--function (or, $L$--function) of this surface.
$L$--function is defined as follows:
\be
L(u)=\prod_{\{\gamma\}}(1-u^{l(\gamma)})^{-1}_{},\label{0.1}
\ee
where the product runs over all primitive (i.e., without rewinding)
closed geodesics on the surface, $l(\gamma)$ being the lengths of
geodesics.

{\bf 3.} In papers \cite{IHARA} an analogue of the Selberg trace
formula was obtained in the case of finite graph with the same, say,
$p+1$, valence for all vertices. Recently in a series of elegant
papers by K.-I. Hashimoto \cite{H-2} and Hyman Bass \cite{B-2} this
relation was generalized to the case of arbitrary finite closed graph:
\be
L(u)=(1-u^2)^{|T|-|E|}\det{}^{-1}(1+u^2Q-uM_1),\label{0.2}
\ee
where $L(u)$ is the corresponding $L$--function (\ref{0.1}) -- an
infinite product over all non--repeated closed paths in the finite
graph, $M_1$ is the operator of summation over all neighbours and $Q$
is a new operator counting valences: $Qx=qx$ if the vertex $x$ has
$q+1$ neighbours. $|T|$ and $|E|$ are total number of vertices and
edges of the graph correspondingly. Note, however, that this operator,
$\Delta(u)$, is neither Laplacian, nor even commutes with $\Delta$
(they coincide only for $u=1$), therefore, straightforward application of
Selberg's ideology in this case  is impossible.

{\bf 4.} The $p$-adic multiloop graphs are non-compact, however, in contrast
to both closed Riemann surfaces and the finite graphs. Nevertheless, there
is a suitable tool for probing in this case. It is the spherical
function technique. Spherical functions are defined as eigenfunctions of the
Laplace--Beltrami operator depending only on distance to a selected point
(the centre). As Laplacian is the second--order operator we always have two
solutions (in a distant point) expanding as $\alpha_+^d$ and $\alpha_-^d$;
$d$ is a distance to the centre and these solutions are interacting only in
the central point. One can find $S$--wave scattering coefficients standing
by these two branches of the solution, and therefrom define Harish--Chandra
$S$--matrix as their ratio. These functions have been found for scattering
on quantum hyperplane \cite{VK} as well as for scattering on $p$-adic
hyperbolic plane \cite{Fr}.
$S$--matrices obtained are closely related to
partition functions of $XXZ$ model and a lot of nice, but somewhat
mysterious, relations between them was obtained in the series of papers by
Freund and Zabrodin \cite{FZ}.

In the present paper we define an analogue of the spherical function for the
$p$-adic multiloop case. It appears that due to non--locality of the central
object whose role the reduced graph $\Tr$ plays now there is a finite set of
possible scenarios of scattering, i{.}e{.}, different possible $S$--matrices
are labelled $c_i$ (the letter ``c'' originates from Harish--Chandra
$c$--function \cite{HC}). When taking the product over all possibilities we
reveal a nice relation:
\be
C=\prod_{i=1}^{|T_{red}|}c_i=\left(\frac{\alpha_+}{\alpha_-}\right)^{|\Tr|}
\,\frac{\det\Delta(\alpha_-)}{\det\Delta(\alpha_+)},\label{0.3}
\ee
where, as before, $\alpha_+$ and $\alpha_-$ are two fixed complex numbers
depending uniquely on the eigenvalue $t$of the Laplacian $D$ acting on the
whole tree $X$: $Df(x)=tf(x)$, and on the initial prime number $p$,
$\alpha_+\alpha_-=1/p$, and $f(x)$ has a general form:
\be
f(x)=A_{ret}(u)\alpha_+^d(x)-A_{adv}(u)\alpha_-^d(x).\label{0.4}
\ee
Here $A_{ret}(u)$, $A_{adv}(u)$ are functions depending only on the
points $u$ of the reduced graph $\Tr$, and $d(x)$ is the distance along the
edges of the
graph $X$ between $x$ and the closest to it point $u$ of $\Tr$. In order for
$f(x)$ to be the spherical function we impose on it the condition
$A_{ret}(u)/A_{adv}(u)=c_i$, and $c_i$ is one and the same for all $u\in\Tr$.

On the contrary, the operator $\Delta(\alpha_{\mp})$ in (\ref{0.3}) is just
the
operator from (\ref{0.2}), $\Delta(u)=1+\t{Q}u^2-u\t M_1$. Here the
operators
$\t Q$ and $\t M_1$ act only on the points of $\Tr$ and we can simply erase
all
branches, that does not alter the form of $\Delta(u)$. Therefore $\Delta(u)$
as
a function depends uniquely on the reduced graph itself and the only
reminiscence of the initial valence of the graph $X$ contains in arguments
$\alpha_+$ and $\alpha_-$ of the function $\Delta(u)$. Comparing (\ref{0.3})
and (\ref{0.2}) we get the relation between Harish--Chandra total
$C$--function
and $L$--function of the $p$-adic curve:
\be
C=\left(\frac{\alpha_+}{\alpha_-}\right)^{|\Tr|}
\left(\frac{1-\alpha_-^2}{1-\alpha_+^2}\right)^{|\Tr|-|E_{red}|}
\,\frac{L(\alpha_+)}{L(\alpha_-)},\label{0.5}
\ee

The idea to consider a proper {\it product\/} of $S$--matrices is originated
from the adelic ideology where we know that in order to compare $p$-adic
quantities with the real ones we are to find a proper product formula. The
one for scattering processes was first proposed in \cite{Fr}, where
amazingly the product of $C$--functions for scattering on $p$-adic
hyperplanes taken over all primes $p$ appeared to be adjusted to
$C$--function of scattering on genus 1 modular figure considered by
L.D.Faddeev and B.S.Pavlov \cite{FP}. The very general formulas concerning
the scattering on symmetrical spaces and gamma--function technique
one can find in \cite{GK}.

We feed the hope to find a proper {\it product\/} of the $L$--functions
appearing in our approach in order to connect them to the ones for geodesics
on Riemann surfaces. We hope that it deserves a big field for forthcoming
activity. The connection of our results to integrable models
alongside results of \cite{FZ} is also to be studied.

The paper is organized as follows: section~1 contains definitions,
interpretation of $p$-adic multiloop curves as graphs, description of
authomorphic functions on these graphs, operators acting in linear spaces of
vertices and oriented edges of graphs together with series of useful
relations connecting these operators. In section~2 action of the Shottky
group on the initial tree is considered and a subtree $D(X)$ is introduced
which is an universal covering of the corresponding reduced graph. Section~3
is devoted to just $L$--functions associated with the group, or,
equivalently, with the reduced graph. The theorem by Hashimoto and Bass is
formulated and the scheme of proof is presented. Eventually, in section~4 we
introduce spherical functions on multiloop graphs, defining simultaneously
corresponding $s$-matrices, and show that the product over all possible
$s$ for a given graph is expressed as a ratio of two $L$--functions.

\newsection{Definitions}
Let $p$ be a natural number and $X=(S(X),\EX)$ be a homogeneous tree
of the order $p+1$, $S(X)$ and $\EX$ being sets of vertices and
edges of the tree correspondingly. We consider a set of {\it
oriented\/} edges, i.e. we can assign one from two opposite orientations
to each edge, therefore two-dimensional subspace corresponds to each edge.
A group $\Pi$ acts freely  on the
tree that means that there is no fixed points inside tree for all
non-unit elements of this group. Each element $\sigma\in\Pi$,
$\sigma\ne 1$ induces a translation of the tree as a whole
along the axis $D(\sigma)$ of the
element. This axis is an infinite line along edges of the tree.
For each two points $x,y$ of the tree there exists a distance
$d(x,y)$ which is equal to the length of the unique way connecting
these two points. The {\it amplitude}
$l(\sigma)=\hbox{inf}\{d(x,\sigma(x))|\ x\in S(X)\}$, where obviously
the minimum is reached on the set $x\in D(\sigma)$. Explicitly an
element $\sigma$ defines a translation on the distance $l(\sigma)$
along the line $D(\sigma)$.

A centralizer $Z(\sigma)$ of the element $\sigma\in\Pi$ is a cyclic
group. If besides it $\sigma$ is a generator of $Z(\sigma)$ then
$\sigma$ is called a primitive element of $\Pi$. It is useful to
define a von Mangolt function $\Lambda(\sigma)=l(\varpi)$, where
$\sigma=\varpi^m$ where $\varpi$ is a primitive element of $\Pi$.

\def\Xr{X_{red}}
The union of all $D(\sigma)$ is some infinite subgraph of the tree
$X$. We call it $D(X)$. It does not necessarily coincide
with the whole tree $X$, therefore a quotient $S(X)/\Pi$ would be
an infinite graph in contrast with \cite{IHARA}. Also the subtree
$D(X)$ is no more uniform in general.

\subsection{Interpretation in terms of graphs}

Let $T=\Pi\backslash X$ be a quotient graph. Then we can release two
essential parts of it: $T=T_{red}\cup B(T)$, where $T_{red}=D(X)/\Pi$
is a finite graph with the number of loops (genus) $g$
coinciding with the number
of generators of the group $\Pi$. This graph also contains no terminal
points, i.e. such points that they are incident to only one edge in $\Tr$.
We call this finite graph the {\it reduced graph}.
$B(T)$ is a set of branches growing
from the vertices of $T_{red}$ in such a manner that the valences of
all vertices of $T$ are the same --- $p+1$. An example of such factorized
tree $T$ for $p=3$ and $g=1$ is presented on Fig.~1.
Note that there is no
restrictions on valences of the vertices inside $T_{red}$ --- all
vertices from 2--valent to $p+1$--valent are possible. There is a
natural notion of closed geodesics for such graphs as well.
For each point $y\in T$ it is useful to define its distance to the
reduced graph $d(y,T_{red})$ as $\inf_{x\in T_{red}}d(x,y)$ which we
also sometimes denote simply as $d(y)$. Note that this minimal distance
is between the point $y$ and a unique point $x\in T_{red}$ which we
shall call an image $t(y)$ of the point $y$. Due to this definition
each branch $B$ is naturally projected into its summit $t\in T_{red}$.


\phantom{xxx}


\begin{picture}(190,2)(-50,85)

\put(100,50){\oval(40,40)[b]}
\put(100,50){\oval(40,40)[t]}
\multiput(100,70)(-20,-20){2}{\line(-1,1){20}}
\multiput(100,70)(20,-20){2}{\line(1,1){20}}
\multiput(60,30)(20,-20){2}{\line(1,1){20}}
\multiput(140,30)(-20,-20){2}{\line(-1,1){20}}
\multiput(100,70)(20,-20){2}{\circle*{3}}
\multiput(80,50)(20,-20){2}{\circle*{3}}
\put(88,82){\line(0,1){8}}
\put(88,82){\line(1,1){8}}
\put(112,82){\line(0,1){8}}
\put(112,82){\line(-1,1){8}}
\put(88,10){\line(0,1){8}}
\put(96,10){\line(-1,1){8}}
\put(112,10){\line(0,1){8}}
\put(104,10){\line(1,1){8}}
\put(68,62){\line(-1,0){8}}
\put(68,62){\line(-1,-1){8}}
\put(68,38){\line(-1,0){8}}
\put(68,38){\line(-1,1){8}}
\put(132,62){\line(1,0){8}}
\put(132,62){\line(1,-1){8}}
\put(132,38){\line(1,0){8}}
\put(132,38){\line(1,1){8}}
\multiput(88,82)(24,0){2}{\circle*{2}}
\multiput(88,18)(24,0){2}{\circle*{2}}
\multiput(68,38)(0,24){2}{\circle*{2}}
\multiput(132,38)(0,24){2}{\circle*{2}}
\multiput(79,97)(6,0){8}{\circle*{2}}
\multiput(79,3)(6,0){8}{\circle*{2}}
\multiput(53,29)(0,6){8}{\circle*{2}}
\multiput(147,29)(0,6){8}{\circle*{2}}

%
\end{picture}

\vspace{5cm}

\centerline{Figure 1.1.  An example of the factorized tree $T$ for
$g=1$, $p=3$.}

\vspace{6pt}

\subsection{Authomorphic functions on $T$.}
Let us fix an unitary representation $(V,\chi)$ of the group $\Pi$. We
denote
$L(X,V)$ the vector space of functions defined on $S(X)$ and taking values
in $V$. We shall consider among all such spaces subspaces $L(\chi,X,V)$
formed by authomorphic functions $F$ such that $F(\sigma
x)=\chi(\sigma)F(x)$
for all $\sigma\in \Pi$ and $x\in S(X)$. If $Y$ is a fundamental domain of
$\Pi$ in $S(X)$ then each $F\in L(\chi,X,V)$ is completely defined by its
values on $Y$.

In what follows we shall consider the trivial representation
$\chi(\sigma)\equiv 1$. We denote
\be
C_0={\Bbb C}^{(X)},\qquad C_1={\Bbb C}^{\EX}
\ee
spaces of ${\Bbb C}$-valued functions defined correspondingly on spaces of
vertices and oriented edges. Note again that two edges, $\ev0$ and
$\ev0^{-1}$ which differ only by orientation are different vectors of
$C_1$  and no relations like $\ev0=-1\cdot \ev0^{-1}$ holds true.

\subsection{Characteristic operators}
Let $M_n,\ n\in{\bf N}$ be a linear operator ($n$th Hecke operator)
acting in the space $C_0$, $M_n:\,C_0\to C_0$:
\be
M_n(F)=\sum_{y\in S(n,x)} F(y) \label{1}
\ee
where the sum runs over all vertices $y$ with $d(x,y)=n$. These $M_n$
constitute a basis in the centre of endomorphism algebra.
Also their multiplication algebra is
\bea
M_1M_n&=&M_{n+1}+pM_{n-1},\quad n\ge2\nonumber\\
M_1M_1&=&M_2+(p+1)M_0,\quad M_0=\hbox{id}.\label{2}
\eea
Let $F$ be a eigenvector of $M_1$ with eigenvalue $t$:
\be
M_1F=tF.
\ee
Then it is also eigenvector of all $M_n$ and
\be
M_nF=S_n(p,t)F\label{3}
\ee
where $S_n(p,t)$ is a system of orthogonal polynomials in $t$ with a
basic relation $tS_n(p,t)=S_{n+1}(p,t)+pS_{n-1}(p,t)$, $S_1=t$,
$S_2=t^2+p+1$.

We also introduce (following \cite{B-2}) a set of operators acting
between two spaces $C_0\leftrightarrow C_1$. First, we have two {\it end
point maps\/} $\d_1$ and $\d_0$ such that $\d_0(e\in \EX)=x_1\in X$ and
$\d_1(e\in \EX)=x_2\in X$ where $x_1$ and $x_2$ are correspondingly
endpoint and starting point of the oriented edge $e$. For $x\in X$ we put
\be
E_j(x)=\{e\in \EX|\d_je=x\},\quad j=0,1.\label{Ej}
\ee
The inverse acting operators, $\sigma_0$ and $\sigma_1$ act from $C_0$
into $C_1$ as follows:
\be
\sigma_j(x\in X)=\sum_{e\in E_j(x)}^{}\,e,\quad j=0,1,
\ee
so that they put into correspondence with the point $x$ a linear
combination of {\it all} edges incident to this vertex taken with the
proper orientation. On the set of vertices it is useful to define the
operator $Q$ counting valences of vertices. If the vertex $x$ has
$q(x)+1$ nearest neighbours, then
\be
Qx = q(x)\cdot x.\label{Q}
\ee
For uniform $p+1$-valent tree the operator $Q$ is equal to the unit
operator times $p$, but for  a general case of a tree $D(X)$ considered
as a universal covering of some finite graph this operator is obviously
non commute with the Hecke operators. For this nonuniform trees Hecke
operators $M_n$ are no more commutative, but it is possible to write down
formulas analogous to (\ref{2}) using the operator $Q$:
\bea
M_1M_n&=&M_{n+1}+QM_{n-1},\quad n\ge2\nonumber\\
M_1M_1&=&M_2+(Q+1)M_0,\quad M_0=\hbox{id}.\label{MQ}
\eea

On the space $C_1$ we define first the inversion map $J$ which simply
change all orientations on edges:
\be
J(\ev{})=\ev{}^{-1},\quad \bigl(\ev{}^{-1}\bigr)^{-1}_{}\equiv\ev.
\label{J}
\ee
Next, we need in an analogue of the set of the Hecke operators for the
space $C_1$. But before doing it we introduce the following useful
notation.

{\bf Paths}. A (oriented) path in $\EX$ is a sequence (finite or
infinite) $c=(\ev1,\dots,\ev{m})$ of edges such that
$\d_0\ev{i}=\d_1\ev{i-1}$, $(1<i\le m)$. We call this path to be of the
length $m$ from the vertex $a=\d_0\ev1$ to $b=\d_1\ev{m}$. We call
$c$ {\it reduced\/} if it does not contain backtrackings, \ie,
$\ev{i}\ne\ev{i-1}^{-1}$ for $1<i\le m$. The path $c$ is {\it closed\/} if
$a=b$. The {\it proper closed path\/} is the closed reduced path with
$\ev1\ne\ev{m}^{-1}$.

Now we are able to define the operator $T$:
\be
T(\ev{0})=\sum_{(\ev{0}\ev1)_{red})}^{} \ev1,\label{T}
\ee
where the sum runs over all possible reduced paths of the length 2.
Treating $T$ as the first Hecke--like operator $T_1$ we can obviously
define $T_n$ as the sum over all oriented reduced paths of the length
$n+1$ of the terminal edges of these paths:
But in contrast to the case
with $C_0$ the relation between $T_n$ and $T_1$ is simply $T_n=T_1^n$ and
the family of these operators is commutative for every tree. (It is an
obvious consequence of the ban for backtrackings on the reduced paths.)
Thus,
\be
T^m(\ev0)=\sum_{(\ev0,\ev1,\dots,\ev{m})_{red}}^{}\,\ev{m}.\label{Tm}
\ee
\def\Du{\Delta(u)}

We introduce now some operator, $\Du$, which will play an important role
in all considerations below. It will appear in various places, so one is
to think that some very useful geometric meaning is encoded in characteristics
of this operator. So, for $u\in {\Bbb C}$, there exists an operator $\Du$
which is well-defined in the sufficiently small vicinity of the origin in
${\Bbb C}$-plane of the variable $u$:
\be
\Du=I-uM_1+u^2Q.\label{Du}
\ee
For $u=1$ $\Du=\Delta$ -- the Laplacian of the graph. We also note that
the operator $\Du$ presents a generating function for the Hecke operators
$M_n$. From the relations (\ref{MQ}) one can derive the following nice
relation concerning $\Du$:
\be
\sum_{n=0}^{\infty}u^nM_n=(1-u^2)\Du^{-1},\label{M.v.D}
\ee
which holds for arbitrary graph.

Next subsection will be denote to an intriguing play in deriving
relations connecting all these operators for arbitrary covering graph.

\subsection{Relations between $M_n$, $\d_j$, $\sigma_j$, $\Du$, $Q$ and
$J$}
We denote $I_0$, $I_1$ unit operators in the spaces $C_0$ and $C_1$
correspondingly.
We start with a series of the relations connecting $\sigma_j$ and $\d_i$:
\bea
&{}&\d_0\sigma_0=\d_1\sigma_1=Q+I_0,\nonumber\\
&{}&\d_1\sigma_0=\d_0\sigma_1=M_1,\label{d-sigma}
\eea
and for the inverse combinations,
\bea
&{}&\sigma_0\d_1=T+J,\nonumber\\
&{}&\sigma_0\d_0=TJ+I_1,\nonumber\\
&{}&\sigma_1\d_1=JT+I_1,\nonumber\\
&{}&\sigma_1\d_0=JTJ+J,\label{sigma-d}
\eea
Put
\bea
\d (u)&=&\d_0u-\d_1,\label{du}\\
\sigma(u)&=&\sigma_0u.\label{su}
\eea
Then,
\bea
\d (u)\sigma(u)&=&\Du -(1-u^2)I_0,\label{dusu}\\
\sigma(u)\d (u)&=&u(T+J)(Ju-I).\label{sudu}
\eea
Now let us consider the linear space $C=C_0\oplus C_1$ and the operators
written in the matrix form acting on this space:
\bea
L&=&\left[\matrix{(1-u^2)I_0 &\d(u) \cr
                       0     & I_1} \right],\nonumber\\
M&=&\left[\matrix{I_0   & -\d(u) \cr
                 \sigma (u) & (1-u^2)I_1} \right].
\eea
The compositions of these two matrices give:
\bea
LM&=&\left[\matrix{ \Du &   0\cr
                   \sigma (u) & (1-u^2)I_1}\right],\label{LM}\\
ML&=&\left[\matrix{ (1-u^2)I_0 &   0\cr
                   \sigma (u)(1-u^2) & (I_1-uT)(I_1-Ju)}\right].\label{ML}
\eea

In what follows we shall need in relation between determinants of
operators $\Du$ on $C_0$ and $I_1-uT$ on $C_1$, and we shall consider the
tree $D(X)$ -- the universal covering of the reduced graph $\Tr$. Since
we consider authomorphic (periodic) functions on the tree, then while
taking the determinants we shall assume that they have to be done for the
reduced graph $\Tr$.
In order to find the relations we need in let us
evaluate determinants for both $LM$ and $ML$ bearing in mind that they
are equal each other.
\be
\det LM=\det\Du (1-u^2)_{}^{2|E_{red}|},
\ee
where $|E_{red}|$ is the number of edges of the finite reduced graph
$\Tr$. For $ML$ we have:
\be
\det ML=(1-u^2)_{}^{|\Tr|}\det (I_1-Tu)\cdot \det (I_1-uJ),
\ee
where $|\Tr|$ is the ``volume'' -- the number of vertices in $\Tr$. The
last thing to do is to find the determinant of $I_1-uJ$. It is
block--diagonal in the basis of edges. Each edge admits two orientations,
therefore
\be
\det (I_1-uJ)=\det \left[\matrix{1 & -u\cr -u & 1}\right]^{|E_{red}|}=
(1-u^2)^{|E_{red}|}_{}.
\ee
Combining all these relations we obtain
\be
\det \Du = \det (I_1-Tu)(1-u^2)_{}^{|\Tr|-|E_{red}|},\label{Du.v.T}
\ee
the value $|\Tr|-|E_{red}|$ is often called Euler characteristic of the
reduced graph $\Tr$.

\newsection{$\Pi$--action on the trees $X$ and $D(X)$}

In this section we remind some facts about the construction of the action of
the finitely generated freely acting group $\Pi$ on the tree $X$.

We adopt the interpretation of the boundary of the tree $X$ as the $p$-adic
projective plane $P_1({\Bbb Q}_p)$~\cite{M-S}. Then the full group of
motions of the tree $X$ is the projective group $PGL(2,{\Bbb Q}_p)$.
The group $\Pi$ is a discrete subgroup of this group and it acts freely that
means that there is no stable points inside the tree for all elements of
this group except the unit one. To each element $\sigma$ of the group we may
put into
correspondence the {\it invariant axis\/} of this element which is the
unique infinite oriented way $\ldots\ev{-1}\ev0\ev1\ev2\ldots$
in the tree that maps to itself under the action of $\sigma$:
$\sigma(\ev{i})=\ev{i+l}$, where $l\equiv l(\sigma)$ is the length of the
element. (We also assume $l(\sigma)<\infty$.)

Let us consider now the subtree $D(X)$ which is, from the one hand, the
universal covering graph for the reduced graph $\Tr$ and, from another hand,
the union of all invariant axes of the elements of $\Pi$.
Of course, while the element $\sigma$ has the length
$l(\sigma)$ it means
that the sequence $\ldots\ev{-1}\ev0\ev1\ev2\ldots$ has the structure
$\ldots(\ev1\ev2\dots\ev{l})(\ev1'\ev2'\dots\ev{l}')
(\ev1''\dots\ev{l}'')\ldots$, where $\ev{i}^{(n)}$ are copies of the edge
$\ev{i}\in \Tr$. Moreover, the sequence $\ev1\ev2\dots\ev{l}$ must be the
proper closed path in $\Tr$ (with possible repetitions, \ie some of $\ev{i}$
may coincide in $\Tr$).

Let us consider the set of conjugate classes $\{\sigma\}$ of the group
$\Pi$:
\be
\{\sigma\}:\left\{\cup_{\sigma, \gamma\in \Pi} \gamma\sigma\gamma^{-1}\in\Pi
\right\}.
\ee
For each element $\beta\in\{\sigma\}$ and each vertex $x\in D(X)$ we have
$\beta x=\gamma\sigma\gamma^{-1}x=y$, hence
$(\gamma^{-1}y)=\sigma(\gamma^{-1}x)$. It means that for each element
$\beta\in\{\sigma\}$ its invariant axis is an {\it image\/} under the action
of some element $\gamma^{-1}$ of the invariant axis of the generating
element $\sigma$. Therefore we established a one--to--one correspondence
between conjugate classes $\sigma$ and a set of all proper closed paths in
the finite graph $\Tr$.

We denote {\it primitive\/} conjugate classes $\{\varpi\}$ such classes
$\{\sigma\}$ that they are generated by the primitive elements of $\Pi$.
Then the proper closed path corresponding to this $\{\varpi\}$ is such
proper closed path in $\Tr$ that it cannot be represented as a power of a
path of smaller length.

\newsection{$L$--function of the factorized tree.}

In this section we consider the graph $D(X)$ which is a
tree graph with the same valence $q(x)+1$ for all images of the
vertex $x\in\Tr$. Let
\be
L(\chi,u)=\prod_{\{\varpi\} }\frac
1{\det(I_V-\chi(\varpi)u^{l(\varpi)})}, \label{1.1}
\ee
where the product runs
over all primitive conjugate classes $\{\varpi\}$ of the group $\Pi$. This
product is absolutely convergent in the domain $\{u:\,
|u|<q_{max}^{-1}\}$, ($q_{max}\le p$ -- the incident number of the
initial tree $X$).
Also we have
\be
\frac d{du}\log \bigl(
L(\chi,u)\bigr)=\sum_{\{\sigma\}}^{}\Lambda(\sigma)
\tr (\chi(\sigma))u^{l(\sigma)-1},\label{1.2}
\ee
where the sum is going over all conjugate classes $\{\sigma\}$ distinct
of $\{1\}$ and the term on the rhs is the logarithmic derivative of
$L(\chi, u)$.

We present now the proof by H.Bass~\cite{B-2} of the main theorem by
K.I.~Hashimoto \cite{H-2}.

\noindent
{\bf Theorem~1}. {\it The function $L(\chi,u)$ is the rational
function of the variable $u$}:
\be
L(\chi,u)=\det(I_1-u\cdot T)^{-1},\label{L.v.T}
\ee
{\it where for the case of the trivial representation $\chi\equiv 1$ $T$ is
defined in} (\ref{T}).

We also denote $Z(\sigma)\equiv L(\chi,u)$ -- zeta-function is $L$-function
for the case of the trivial representation.

 From (\ref{Du.v.T}) we immediately get

\noindent
{\bf Corollary~1}. {\it In above notations,\/}
\be
L(\chi=1,u) = \det \Du^{-1}(1-u^2)_{}^{|\Tr|-|E_{red}|}.\label{L.v.Du}
\ee

Let us choose in the tree $D(X)$ some finite {\it fundamental domain\/}
$F(\Pi)$ of the symmetry group $\Pi$. Our aim is now to consider the
contribution to the trace of the operator $T^m$ from some edge $\ev1\in
F(\Pi)$. Only those $\ev{i}\in D(X)$ give the contribution to $\tr T^m$
which
\begin{enumerate}
\item lie on the distance $m$ to the initial edge $\ev1$ (along some reduced
path $\ev1\ev2\dots\ev{m+1}$);
\item have the orientation along this path.
\end{enumerate}
The crucial observation follows. Let us consider the action of some element
$\sigma\in\Pi$ on oriented edges of $D(X)$ (Fig.~2). It is easy to see that
as soon as the edge does not belong to the invariant axis $D(\sigma)$ of this
element (like $\ev{x}$ does on Fig.~2) then the orientation of the image of
it $\sigma\ev{x}$ is {\it opposite\/} to the orientation of the reduced path
approaching from $\ev{x}$ to $\sigma\ev{x}$, thus such edges do not
contribute to the trace! And only edges which like $\ev1$ belong to the
invariant axis $D(\sigma)$ preserve their orientation toward $D(\sigma)$ when
are undergone to the translation (like $\ev{l+1}$ is $\sigma\ev1$ on Fig.~2).


\phantom{xxx}
\begin{picture}(190,2)(0,85)

\multiput(20,20)(25,0){7}{\vector(1,0){25}}
\multiput(45,20)(125,0){2}{\circle*{2}}
\put(195,20){\line(1,0){50}}
\multiput(45,20)(125,0){2}{\line(-1,2){30}}
\multiput(45,20)(125,0){2}{\line(1,2){30}}
\multiput(15,80)(125,0){2}{\circle*{2}}
\multiput(15,80)(125,0){2}{\vector(1,-2){10}}
\multiput(170,20)(-10,20){2}{\vector(-1,2){10}}
\put(50,10){\makebox{$\ev1$}}
\put(175,10){\makebox{$\ev{l+1}\!=\!\sigma\ev1$}}
\put(230,10){\makebox{$D(\sigma)$}}
\put(20,75){\makebox{$\ev{x}$}}
\put(145,75){\makebox{$\sigma\ev{x}$}}
\put(155,55){\makebox{$T^m\ev1$}}

%
\end{picture}

\vspace{4.5cm}

\centerline{Figure 2.  The action of the group element $\sigma\in\Pi$ on the
edges of $D(X)$ (or $X$ itself),}
\centerline{with the length $l(\sigma)=l$. Note that $\sigma\ev{x}\ne
T^{m+1}\ev1$ for all $m$.}

\vspace{5pt}

{\it Note\/}. It is this point where the structure of $\tr T^m$
essentially differs from the structure of, say, $M_m$ -- the $m$th Hecke
operator for the space $C_0$ since for $\tr M_m$ all pairs of vertices
$x,y\in D(X)$ and not only
those which belong to the invariant axis of an element $\sigma$ have to
contribute to this trace as far as $d(x,y)=m$ and $y=\sigma x$.

Now let us return to the space $C_1$. Now we fix for a moment the edge
$\ev1\in D(X)$.
As we have mentioned already, to each
primitive conjugate class $\varpi$ it corresponds not necessarily one axis
$D(\varpi)\subset D(X)$ such that $\ev1\in D(\varpi)$. We denote
$\av{i}$  the oriented edges of the reduced graph $\Tr$ itself, thus each
$\ev{i}\in D(X)$ is an image of the vector $\av{i}\in \Tr$ with orientation
naturally preserved. Then for each element $\sigma\in \Pi$ we may put into
correspondence the sequence of ``letters''
\be
D(\sigma)\simeq \ldots(\av1\av2\dots\av{l})(\av1
\av2\dots\av{l})\ldots, \label{D(sigma)}
\ee
where $l\equiv\Lambda(\sigma)=l(\varpi)$ -- the
length of the generating element for $\sigma$. It is clear that if
$l(\sigma)=m$, then $l(\varpi)=l$, $l|m$ that is $l$ is a divisor of $m$.
Note again that there can be repeated $\av{i}$ in the cyclic sequence
$\av1\av2\dots\av{l}(\av{l+1}\equiv\av1)$,
Moreover, this sequence may contain subperiods if $\sigma$ is not a
primitive element, but obviously the minimal length of this subperiod, $l$,
is a divisor, $l|m$.

Now let us fix $\av1$ -- the pre-image of $\ev1$. When we are interested in
its contribution to $\tr T^m$ we are to find all possible different cyclic
expressions
\be
\av1\av2\dots\av{m}(\av{m+1}=\av1). \label{av1--avm}
\ee
Now we are going to
establish a correspondence between these two sets, (\ref{D(sigma)}) and
(\ref{av1--avm}).

{\bf 1.} To each finite cyclic sequence $\av1\dots\av{m}(\av1)$ we may
inambiguously find an infinite periodic sequence $\dots(\av1\dots\av{m})
(\av1\dots\av{m})\dots$ corresponding to unique element of primitive
conjugate class $\{\varpi\}$ with $l(\varpi)=l|m$.

{\bf 2.} On the contrary, choose an element from $\{\varpi\}$, or,
equivalently, some arbitrary periodic reduced sequence
\be
\ldots(\dots)(\av{i_1}\av{i_2}\dots\av{i_l})(\av{i_1}\dots)\dots
\label{p1}
\ee
with no subperiods and the minimal period $l=l(\varpi)$. If it contains
the edge $\av1$ (together with orientation) $d_1$ times among the edges
$\{\av{i_1},\dots,\av{i_l}\}$ and, moreover, $l$ is the divisor of $m$, then
there are exactly $d_1$ {\it different\/} sequences $\av1\dots\av{m}(\av1)$
containing in (\ref{p1}). Eventually, doing the sum over all edges of $\Tr$
(or of the fundamental domain $F(\Pi)\in D(X)$), \ie evaluating the trace of
the operator $T^m$ we find that the contribution from the sequence
$\ldots(\dots)(\av{i_1}\dots\av{i_l})\dots$ to this trace is equal to
\be
\sum_{j=1}^{2|E_{red}|}\#\left\{\av{j}\ \hbox{in}\
\{\av{i_1},\dots,\av{i_l}\}\right\}\equiv l,
\ee
$l$ being the total length of the primitive element $\varpi$. Therefore we
have the following remarkable formula basic for our consideration:
\be
\tr T^m=\sum_{l|m}^{}l\cdot \#\bigl\{\varpi:l(\varpi)=l\bigr\},
\label{T1m}
\ee
where the sum runs over all primitive conjugate classes of $\Pi$.
Now we have from (\ref{1.2}) for zeta-function $Z(\sigma)=L(\chi\!=\!1,u)$:
\bea
u\frac d{du}\log \bigl(
L(\chi\!=\! 1,u)\bigr)&=&\sum_{\{\varpi\}}^{}\frac{l(\varpi)\cdot
u^{l(\varpi)}}{1-u^{l(\varpi)}}=
\sum_{\{\varpi\}}^{}\sum_{n=1}^{\infty}l(\varpi)\cdot
u^{nl(\varpi)}\nonumber\\
&=&\hbox{from\ (\ref{T1m})}\ \sum_{k=1}^{\infty}\tr T^k\cdot u^k
=-u\frac{d}{du}\log\,\det(1-u\cdot T).\label{LT}
\eea

Therefore we have proved the Theorem~1 by K.-I.~Hashimoto and
H.~Bass (\ref{L.v.T}).

\newsection{Spherical functions on factorized trees.}

In this section we consider spherical functions on the factorized tree
$T$. First, we shall consider it on the tree $S(X)$ itself. Choose the
point
$X\in S(X)$ and claim it the center of the tree. Then the spherical
function is
an eigenvector of $M_1$: $M_1F(n,x)=tF(n,x)$ with the dependence only on the
distance from the point $x$, i.e. it is constant on each sphere $S(n,x)$.
This problem can be easily solved and the answer is the following (for the
trivial representation $\chi\equiv I$):
\be
F(n,x)=a_+\alpha_+^n-a_-\alpha_-^n,\label{4}
\ee
where $\alpha_{\pm}=\frac{t}{2p}\pm\sqrt{\frac{t^2}{4p^2}-\frac 1p}$,
$\alpha_+\alpha_-=1/p$,
\be
a_+=\frac{p\alpha_+-\alpha_-}{(p+1)(\alpha_+-\alpha_-)},\quad
a_-=\frac{p\alpha_--\alpha_+}{(p+1)(\alpha_+-\alpha_-)}.\label{5}
\ee

Now we are to define the same object for a general factorized graph $T$. In
order to do it we shall consider a superposition of the solutions (\ref{4})
with the sources $x$ placed in the vertices of $D(X)$:
\be
F(x)=\sum_{y\in D(X)}s_yF(d(x,y),y).\label{5a}
\ee
(Remind that $D(X)$
is the universal covering of the {\it reduced graph} $T_{red}$.) In order
to
be well defined on $T$ it is necessary to put the function $s_y$
periodic under the action of $\Pi$: $s_{\sigma y}=s_y$ for all
$\sigma\in\Pi$ and $y\in D(X)$. We place the sources {\it only\/} at the
points of $D(X)$ in order to ensure that the behaviour of this solution on
each branch growing from $\Tr$ is given by (\ref{4}) with common factor
depending only on the point of $\Tr$ from which this branch is growing.

Let us introduce a kern function $K$:
\be
K(z,y|x)=\sum_{\sigma\in\Pi}x^{d(z,\sigma(y))},\label{6}
\ee
where $z,y$ are points of $D(X)$. In fact, this function is periodic under
the action of $\Pi$ over both its arguments separately. Therefore we can
think about $z$ and $y$ that they are points on reduced graph $\Tr$ itself.
This function is also symmetric in $z$ and $y$.
Then the function (\ref{5a}) acquire the form:
\be
F(x)=\sum_{y\in T_{red}}s_y\bigl[a_+K(z(x),y|\alpha_+)\alpha_+^{d(x)}-
a_-K(z(x),y|\alpha_-)\alpha_-^{d(x)}\bigr],\label{7}
\ee
where $z(x)$ is the closest to $x$ point of the reduced graph. (If $x$
itself is a point of $\Tr$, then $z(x)=x$.)
We shall treat one part of the expression proportional to $\alpha_+^{d(x)}$
as a retarded wave function and other part --- as an advanced wave function.
So $F(x)$ has a general form
\be
F(x)=A_{ret}(t(x))\alpha_+^{d(x)}-
A_{adv}(t(x))\alpha_-^{d(x)}.
\ee
Up to the present moment we didn't fix anyway the coefficients $s_y$. Now
it is a proper moment to do it.

{\bf Definition.} A spherical function $F(x)$ is an eigenfunction of the
operator $M_1$ of the form (\ref{7}) such that the {\it ratio} of the
coefficients $A_{ret}(t(x))/A_{adv}(t(x))$ is constant for all $x\in T$.
We call this constant $c$.

It means that $c$ is a solution to the following linear equation system:
\be
\sum_{y\in T_{red}}s_ya_+K(z,y|\alpha_+)=c\sum_{y\in
T_{red}}s_ya_-K(z,y|\alpha_-).
\label{8}
\ee
Obviously it has as many solutions as the number $|T_{red}|$ of vertices in
$T_{red}$.  We do not intend to solve all these equations but one
characteristic is of the great interest, namely, the product over all
possible
solutions $c_i$:
\be
C=\prod_{i=1}^{|T_{red}|}c_i.\label{9}
\ee
This $C$ --- the {\it total\/} $S$--matrix, is a direct generalization of
the Harish--Chandra $c$--function to the case of the multiloop $p$-adic
curves. From the
equation (\ref{8}) it is clear that
\be
C={\det K(z,y|\alpha_+)\over \det K(z,y|\alpha_-)}\,
\left(\frac{a_+}{a_-}\right)^{|T_{red}|},\label{10}
\ee
where the determinant is taken over $|T_{red}|\times
|T_{red}|$ matrices parameterized by the points of $T_{red}$.

\subsection{Determinant of the operator $K(z,y|\alpha)$.}

In what follows we deal with the trivial representation of the group
$\Pi$ $\chi(\sigma)\equiv 1$.

\def\Kt{{\widetilde{K}}}
In this technical subsection we find an explicit form of  $\det
K(z,t|x)=\e^{\tr\log K}$. First we release trivial ways from the sum
(\ref{6}):
\be
K(z,y|x)=\delta_{\Pi}(z,y)+\Kt(z,y|x),\label{2.1}
\ee
where
\be
\delta_{\Pi}(z,y)=\left\{ \begin{array}{ll} 1, & z=y\ \  z,y\in
T_{red},\\ 0,& \mbox{otherwise}\end{array} \right.
\ee
is the unit operator.

Now let us find a operator $\Delta_z$ for which the function $K(z,y|x)$
is a Green function:
\be
\Delta_zK(z,y|x)=\delta_{\Pi}(z,y),\label{2.2}
\ee
and, obviously $\det K=\det\Delta_z^{-1}$. We also need to define
analogies of the operator $\t M_1$, $\t Q$ and the sphere $\t S(n,z)$ for the
points of the reduced graph. We define for $z\in D(X)$:
\bea
\t S(n,z)&=&\{z_i\in D(X):\, d(z,z_i)=n\}, \nonumber\\
\t Qf(z)&=&q(z)f(z), \ \hbox{where}\ q(z)=\#\{z_i\in \t S(1,z)\}-1,
\nonumber\\
\t M_1(f(z))&=&\sum_{z_i\in S^r(1,z)}f(z_i).\label{2.3}
\eea

We look for the operator $\Delta_z$ in the form
\be
\Delta_z f(z) \stackrel{\rm def}{=}a(z)\t M_1f(z)+b(z)f(z),
\label{2.4}
\ee
where $z\in T_{red}$.
\bea
\Bigl.\Delta_z K(z,y|x)\Bigr|_{z\ne y}
&=&a(z)\sum_{\sigma\in\Pi}^{}\sum_{t\in S(1,\sigma(z)}^{}x^{d(y,t)}
+b(z)\sum_{\sigma\in\Pi}^{}x^{d(\sigma(z),y)}\nonumber\\
&=&a(z)\sum_{\sigma\in\Pi}^{}x^{d(\sigma(z),y)}\left[q(z)x+\frac
1x\right]+ b(z)\sum_{\sigma\in\Pi}^{}x^{d(\sigma(z),y)}=0.
\label{2.5}
\eea
Here $q(z)+1$ are  valences of points $z\in T_{red}$. In the formula
(\ref{2.5}) we essentially used the fact that among the points $t\in
S(1,\sigma(z))\subset D(X)$ there are exactly $q(z)$ points which
belong to $S\bigl(d(\sigma(z),y)+1,y\bigr)$ and the unique point
belonging to $S\bigl(d(\sigma(z),y)-1,y\bigr)$. From (\ref{2.5}) we
get the first restriction on $a(z)$, $b(z)$:
\be
a(z)\left[q(z)x+\frac 1x\right]+b(z)=0.\label{2.6}
\ee
Eventually, for the case of coinciding points $z=y$ we have
\bea
\lefteqn{\Bigl.\Delta_z K(z,y|x)\Bigr|_{z=y}
=a(z)\sum_{\sigma\in\Pi}^{}\sum_{t\in S(1,\sigma(z)}^{}x^{d(z,t)}
+b(z)\sum_{\sigma\in\Pi}^{}x^{d(\sigma(z),z)}}\nonumber\\
&{}&=\sum_{{\sigma\in\Pi\atop
\sigma\ne I}}^{}x^{d(\sigma(z),z)}\left[a(z)\left(
q(z)x+\frac 1x\right)+b(z)\right]+ a(z)[q(z)+1]+b(z)=1.
\label{2.7}
\eea
 From here we have the second equation on $a(z)$, $b(z)$:
\be
b(z)(q(z)+1)+a(z)=1.\label{2.8}
\ee
The solution to the equations (\ref{2.6}), (\ref{2.8}) has the form:
\be
a(z)=-\frac{x}{1-x^2},\quad\quad b(z)=\frac{1+q(z)x^2}{1-x^2}.
\label{2.9}
\ee

Therefore, the function $K(z,y|u)$ is a kernel of the operator
$(1-u^2)\Du${}\footnote{Therefore, we just have proved the formula
(\ref{M.v.D})}.
For the determinant of $K(z,y|x)$ we have
\be
\det K(z,y|x)=\det{}^{-1}\Bigl[(1+q(z)x^2)I-xM^r_1\Bigr]
(1-x^2)^{|T_{red}|}.
\label{2.10}
\ee

Comparing this formula with (\ref{L.v.Du}) for the arbitrary
reduced graph, we establish a relation for $L$-function, or, equivalently,
$p$-adic $Z$-function which is $L$-function for the trivial representation:
\be
Z(u)\equiv L(1,u)=(1-u^2)^{-|E_{red}|}\det K(z,y|u),
\label{2.11}
\ee
where $|E_{red}|$ is the number of edges of the reduced graph.

Eventually, for Harish--Chandra function $C$ (\ref{10}) we have
\be
C={\det \Delta(\alpha_-)\over \det \Delta(\alpha_+)}\,
\left(\frac{\alpha_+}{\alpha_-}\right)^{|T_{red}|},\label{4.22}
\ee
and using the fact that $\alpha_-\alpha_+=1/p$, the combination
$(\alpha_\pm)^{|\Tr|}\det\Delta(\alpha_\mp)$ acquires the form
\be
(\alpha_\pm)^{|\Tr|}\det\Delta(\alpha_\mp)=
\det\bigl(\alpha_\pm\Delta((\alpha_\mp)\bigr)
=\det\left(\alpha_\pm+\frac{\alpha_\mp}{p}\t Q-\frac 1p\t M_1\right).
\label{4.23}
\ee

\subsection{Example~1. Graph $T$ with no external branches}
Here we consider a limiting case of our construction when the graph $T$
exactly coincides with $\Tr$. In this no--scattering case $q(z)\equiv p$ and
then the inversion relation for $\Du$ fulfills. If $\alpha_-\alpha_+=1/p$,
we have
\bea
\det{}^{-1}\bigl[(1+p\alpha_+^2)I-\alpha_+M_1\bigr]&=&
\det{}^{-1}\left[1+\frac{1}{p\alpha_-^2}-\frac{1}{p\alpha_-}M_1\right]
\nonumber\\
&=&\bigl(\alpha_-^2p\bigr)^{|\Tr|}
\det{}^{-1}\bigl[(1+p\alpha_-^2)I-\alpha_-M_1\bigr]\nonumber\\
&=&\frac{\alpha_-^{|\Tr|}}{\alpha_+^{|\Tr|}}
\det{}^{-1}\bigl[(1+p\alpha_-^2)I-\alpha_-M_1\bigr],\label{inverse}
\eea
and for the total scattering matrix $C$ (\ref{10}) we get using (\ref{4.22})
and (\ref{inverse}) that $C\equiv 1$.
Thus in this case $C$ is trivial and does not depend at all on the shape of
the reduced (or, in this case, total) graph $\Tr$.

\subsection{ Example~2. A one--loop case ($g$=1).}
Now let us consider the simplest case of the group $\Pi_1$ with the
only generator $\gamma$: each element $\sigma\in \Pi_1$ is presented
in the form $\sigma=\gamma^m$, $m\in {\bf Z}$. The amplitude of
$\sigma$ is then equal to $|m\cdot l(\gamma)|$. In its turn
$l(\gamma)$ is the length $n$ of the ring being a reduced graph
$T_{red}^{(1)}$ for this problem.  As for the function $\Kt_1(z,y|x)$
we present it only in the case of the coinciding arguments:
\be
\Kt_1(z,z|x)=\sum_{m\in{\bf Z},\ m\ne0}x^{|m|n}=2\frac{x^n}{1-x^n}.
\label{12.6}
\ee
It does not depend on the point $z\in T^1_{red}$ and the determinant
can be easily calculated. We shall do it in two ways: directly from
the definition of the operator $\Delta_{(g=1)}$ (\ref{2.4}) and using
formulae for $L$--function (\ref{L.v.Du}), {\sl Theorem~1}. We begin
with direct calculation of the determinant.

Let $\{f_i,\, 1\leq i\leq n\}$ are cyclically arranged
$(f_{n+1}\equiv f_1)$ points of the reduced graph, $M_1^r(f_i)=
f_{i-1}+f_{i+1}$. The operator $\Delta$ has the form:
\be
\Delta f_i=\frac{x}{x^2-1}(f_{i+1}+f_{i-1})-\frac{x^2+1}{x^2-1}f_i,
\label{delta1}
\ee
and we are looking for its eigenvectors ${f_k^{(s)}}=\e ^{ik\cdot
2\pi s/n}$. Then the corresponding eigenvalue $\lambda_s$ is
\be
\lambda_s=b+2a\cos 2\pi s/n,\quad b=-\frac{x^2+1}{x^2-1},\quad
a=\frac x{x^2-1}.
\ee
Doing a product over all $n$ eigenvalues we obtain after a little algebra:
\be
\det \Delta_{(g=1)}=\prod_{s=1}^n \lambda_s=\frac {(1-x^n)^2}{(1-x^2)^n}.
\label{det1}
\ee
Comparing this answer with (\ref{L.v.Du}), (\ref{2.11}) and (\ref{1.1})
we find for the $L$--function (in the case of the trivial character
$\chi(u)$):
\be
L(1,u)_{g=1}=\frac 1{(1-u^n)^2},\label{L1}
\ee
since there are two primitive elements in the group $\Pi_1$, namely
$\gamma$ and $\gamma^{-1}$ with the same length
$l(\gamma)=l(\gamma^{-1})=n$.

\newsection{Conclusion}
Here I want briefly discuss possible prospects of this action. First,
the formula (\ref{4.22}) enables us to investigate a spectrum of the
Laplace operator for a graph. It appears that all necessary information
concerning this spectrum is already provided when considering spherical
functions of discussed type. Another question is to
connect all above mentioned to
integrable systems. Third, one can try to apply this scheme to exploration
of continuous spaces of constant negative curvature in the hope to find
adelic formulas for the product of the results already obtained in $p$-adic
and in continuous cases. The work on all the three topics is in progress
\cite{CERZ}.

\newsection{Acknowledgements}
This work was started during my visit in Enrico Fermi Institute, University
of Chicago, in 1991 due to a lot of stimulating discussions with P.Freund to
whom I am especially grateful. However that time we did not know about the
papers of Hashimoto and Bass and it was A.Nikitin from St.Petersburg
Branch of Steklov Math. Institute who had turned our attention to these
papers that resumed the deadlock in this action. I am greatly indebted to
B.Enriquez and V.Rubtsov for a lot of useful discussions both in Ecol\'e
Polytechnique, Paris, and in Moscow.